\documentstyle[prl,aps,floats,epsf]{revtex}

\begin{document}

\draft

\twocolumn[\hsize\textwidth\columnwidth\hsize\csname @twocolumnfalse\endcsname
\title
{
Classical Correlation-Length Exponent in Non-Universal Quantum
Phase Transition of Diluted Heisenberg Antiferromagnet
}

\author
{ 
C.~Yasuda$^{1}$, S.~Todo$^{1}$, K.~Harada$^{2}$, N.~Kawashima$^{3}$, 
S.~Miyashita$^{4}$, and H.~Takayama$^{1}$ 
}

\address 
{
$^{1}$Institute for Solid State Physics, University of Tokyo, Kashiwa 277-8581, Japan \\
$^{2}$Department of Applied Analysis and Complex Dynamical Systems, Kyoto University, Kyoto 606-8501, Japan \\
$^{3}$Department of Physics, Tokyo Metropolitan University, Tokyo 192-0397, Japan \\
$^{4}$Department of Applied Physics, University of Tokyo, Tokyo 113-8656, Japan
}

\date{\today}

\maketitle

\widetext

\begin{abstract}
Critical behavior of the quantum phase transition of a site-diluted
Heisenberg antiferromagnet on a square lattice is investigated by means
of the quantum Monte Carlo simulation with the continuous-imaginary-time
loop algorithm.  Although the staggered spin correlation function decays 
in a power law with the exponent definitely depending on the spin size 
$S$, the correlation-length exponent is classical, i.e., 
$\nu=4/3$. This implies that the length scale characterizing the 
non-universal quantum phase transition is nothing but the mean size of 
connected spin clusters. 
\end{abstract}

]

\narrowtext

Ground-state phase transitions in two-dimensional (2D) diluted quantum
Heisenberg antiferromagnets (HAF's) have attracted much interests
because they are caused by the coexistence of quantum fluctuations and
randomness\cite{breed,jongh,chakra,cheong,ting,clarke,corti}. Numerical
works\cite{behre,sand} and theoretical works\cite{yasuda,chen} have
given various estimates of the critical concentration of the system with
$S=1/2$. All of them are above the purely-geometrical percolation
threshold on a square
lattice, $p_{\rm cl}$=0.5927460(5)\cite{ziff}. This suggests that the
phase transition could be seriously affected by quantum fluctuations. 

Recently, Kato {\it et al}.\cite{kato}
have investigated the diluted 2D HAF with
$S=$ 1/2, 1, 3/2, and 2 by means of the quantum Monte Carlo
(QMC) method with the continuous-imaginary-time loop 
algorithm\cite{evertz1,evertz2,harada,todo}. Their conclusion is
qualitatively different from the above mentioned results: the critical
concentration coincides with $p_{\rm cl}$ and does not depend on
$S$. The coincidence of the critical concentration with $p_{\rm cl}$ has
also been reported on the bond-diluted HAF\cite{sand2}.

The critical exponents of the phase transition at $p_{\rm cl}$ have also 
been estimated. Kato {\it et al}. have obtained the critical exponent
$\beta$ of the zero-temperature staggered magnetization. Interestingly,
the value of $\beta$ is different from that of the classical
($S=\infty$) exponent and depends on $S$. They have also estimated other 
critical exponents by the finite-size scaling (FSS) analysis assuming
the following form for the static staggered structure factor at zero
temperature:
\begin{equation}
  \label{fss}
  S_{\rm s}(L,0,p)\sim L^{\Psi}\tilde{S}_{\rm s}(L^{1/\nu}(p-p_{\rm cl})) \ ,
\end{equation}
where $L$ is the system size and
\begin{equation}
   \label{str}
   S_{\rm s}(L,T,p)\equiv \frac{1}{L^{d}}\sum_{i,j}{\rm e}^{{\rm i}{\vec k}\cdot 
                ({\vec r}_{i}-{\vec r}_{j})}
                \langle S_{i}^{z}S_{j}^{z}\rangle 
\end{equation}
with ${\vec k}=(\pi, \pi)$ and $d=2$. The bracket $\langle \cdots
\rangle$ in Eq. (\ref{str}) denotes both the thermal and random
averages. The scaling function $\tilde{S}_{\rm s}(x)$ in Eq. (\ref{fss})
has the following asymptotic form:
\begin{eqnarray}
 \label{sf}
 \tilde{S}_{\rm s}(x) 
      &\sim&  \left\{
          \begin{array}{@{\,}ll}
 x^{2\beta}                       & {\rm for}~~x\gg 1 \\
 |x|^{-\nu\Psi}                   & {\rm for}~~x\ll -1 \ ,
                    \end{array}
                  \right. 
\end{eqnarray}
where the exponent $\beta$ is related to $\Psi$ and $\nu$ by the scaling 
relation
\begin{equation}
  \label{relate}
  2\beta = (d-\Psi)\nu \ .
\end{equation}
They have estimated the critical exponent $\Psi$ by the FSS analysis
exactly at $p=p_{\rm cl}$ and have found that it depends on $S$.

Kato {\it et al}. have attributed the $S$-dependence in exponent $\beta$ or
$\Psi$ to quantum fluctuations, while the length exponent $\nu$ is
assumed to be given by the classical one, $\nu_{\rm cl}=4/3$\cite{stau}, which
governs a power-law divergence of the mean size of connected spin clusters as 
$\lambda(p) \propto |p-p_{\rm cl}|^{-\nu_{\rm cl}}$. Namely, they have
assumed that
the staggered spin correlation function between two spins in a cluster
is described by the scaling expression 
\begin{equation}
     \label{corr2}
  C(i,j;p) \sim  r_{i,j}^{-\alpha} \tilde{C}(r_{i,j}/\lambda(p)) \
\end{equation}
with $\tilde{C}(x) \sim$ const. at $x \ll 1$. The power-law decay of the
correlation function is due to quantum fluctuations and its
$S$-dependent exponent $\alpha$ is related to $\Psi$ by
\begin{equation}
     \label{relate2}
  \Psi = 2D -d - \alpha,
\end{equation}
where $D$ is the fractal dimension ($91/48$ for $d=2$). Kato {\it et
al}. have checked
this scenario simply by evaluating $\nu$ through Eq. (\ref{relate})
using $\beta$ and $\Psi$ obtained by their simulation. 

In the present paper, we perform the FSS analysis on $S_{\rm s}(L,0,p)$
of systems with $S=1/2$ and 1 more systematically by carrying out the QMC
simulation at various concentrations not only at $p_{\rm cl}$. It is
found, as we see below, that our QMC data are well described by the FSS form
of Eq. (\ref{fss}) in a whole range of $p$ studied, which includes the 
asymptotic ranges $x\gg 1$ and $x\ll -1$ in Eq. (\ref{sf}). Within
numerical accuracy of our analysis, exponent $\Psi$ turns out to
definitely depend on the spin size $S$, while exponent $\nu$ does
not. The latter $S$-independent value of $\nu$ coincides with the
classical one (=4/3) within the error bar. These results further support
the above mentioned scenario of the quantum phase transition in the diluted 2D
HAF, particularly, the ansatz that the length scale characterizing the
transition is nothing but the mean size of connected spin clusters.

The system we study is the site-diluted HAF on a square lattice
described by the Hamiltonian
\begin{equation}
    {\cal H} = J\sum_{\langle i,j \rangle} \epsilon_{i}\epsilon_{j}
              {\bf S}_{i} \cdot {\bf S}_{j} \ ,
\end{equation}
where $J(>0)$ is the antiferromagnetic coupling constant, 
$\sum_{\langle i,j \rangle}$ denotes the summation over all
nearest-neighbor pairs and ${\bf S}_{i}$ is the quantum spin operator at
site $i$. The quenched magnetic occupation factors $\{\epsilon_{i}\}$
independently take 1 or 0 with probability $p$ and $1-p$, respectively.
We simulate $L\times L$ square lattices with the periodic boundary
condition by means of the same QMC method with the
continuous-imaginary-time loop algorithm as that adopted by Kato {\it et 
al}. An improved estimator is used to calculate the static
staggered structure factor.  At each parameter set ($L,T,p$) physical
quantities of interest are averaged over 1000 -- 3000 samples.
At each sample, 10$^{3}$ -- 10$^{4}$ Monte Carlo steps (MCS) are spent
for measurement after 500 -- 10$^{3}$ MCS for thermalization. 

In a finite system, $S_{\rm s}(L,T,p)$ converges rapidly to its
zero-temperature values at temperatures lower than the gap due to the
finiteness of the system. 
The saturation temperature turns out to be smaller for smaller $|p-p_{\rm cl}|$. 
In the present work $S_{\rm s}(L,0,p)$ close to
$p_{\rm cl}$ is approximated by $S_{\rm s}(L,T,p)$ at the following temperatures: for
the $S=1/2$ system $T=0.002J$ for $L=24$ and $T=0.001J$ for $L=32$, 40,
and 48, and for the $S=1$ system $T=0.01J$ for $L=24$ and $T=0.005J$ for 
$L=32$, 40, and 48. 

\begin{figure}[tb]
 \epsfxsize=0.47\textwidth
 \epsfbox{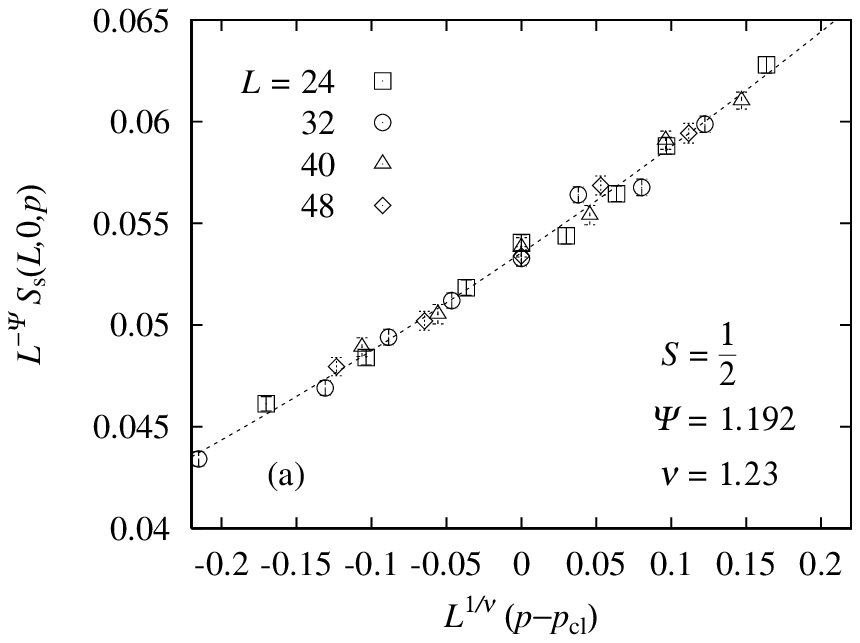}
 \vspace*{0.5em}\hspace*{0em}

 \epsfxsize=0.47\textwidth
 \epsfbox{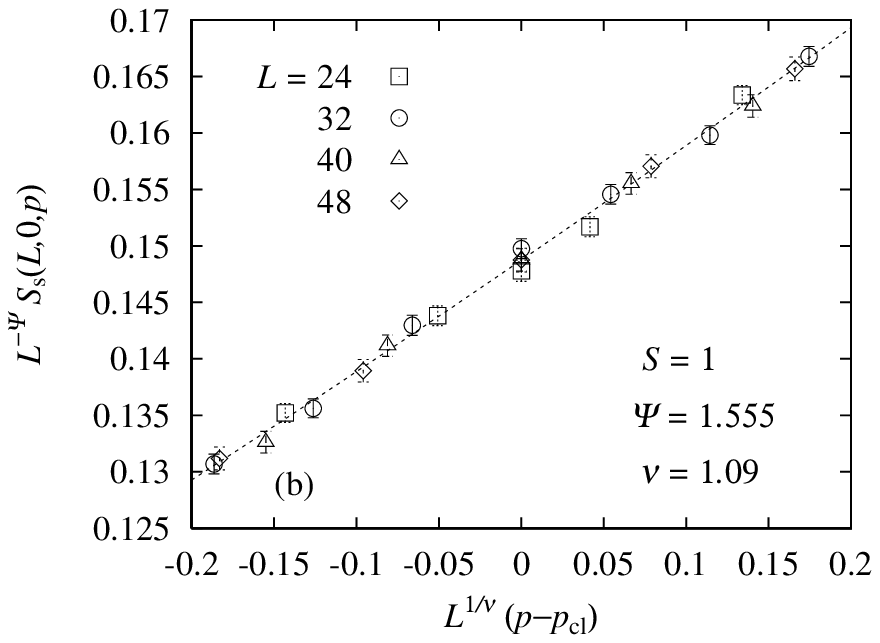}
 \vspace*{0.5em}\hspace*{0em}
 
 \caption{Scaling plot of $S_{\rm s}(L,0,p)$ for (a) $S=1/2$ and (b)
 $S=1$. The dashed line represents $\tilde{S}_{\rm s}(x)$ which is approximated
 by a polynomial of order 2.}
\end{figure}

Let us first examine the FSS analysis of $S_{\rm s}(L,0,p)$ at $p$ close
to $p_{\rm cl}$. The results of the $S=1/2\ (S=1)$ system at 
$0.580 \leq p \leq 0.605$ ($0.585 \leq p \leq 0.600$) are shown in
Fig. 1(a) (1(b)). The error bars in the figures represent the standard
deviation. In the FSS fit the critical
concentration is
set to be $p_{\rm cl}$ ($\simeq 0.5927460$), and the scaling function 
$\tilde{S}_{\rm s}(x)$ is approximated by a polynomial of order 2. As
seen in Fig. 1(a), the QMC data for $S=1/2$ are well scaled with 
$\Psi=1.192$ and $\nu=1.23$. The statistical accuracy of the fit is
shown in Fig. 2, where we draw the confidence region within which 
the true values of $\Psi$ and $\nu$ fall with probability 68.3\%
(1-$\sigma$), 95.3\% (2-$\sigma$), or 99.7\% (3-$\sigma$). Similarly, the data
for $S=1$ are well scaled with $\Psi=1.555$ and $\nu=1.09$ as seen in 
Figs. 1(b) and 2. From these results we can conclude that exponent 
$\Psi$ definitely depends on the spin size $S$. The exponents $\nu$ of
$S=1/2$ and 1, on the other hand, coincide with each other and with 
its classical value ($=4/3$) within the numerical accuracy of the
present analysis. The obtained values of $\Psi$, $\nu$, $\alpha$, and
$\beta$ are summarized in Table I. The exponents $\alpha$ and $\beta$
are calculated by the scaling relations, Eqs. (\ref{relate}) and 
(\ref{relate2}).

\begin{figure}[tb]
 \epsfxsize=0.47\textwidth
 \epsfbox{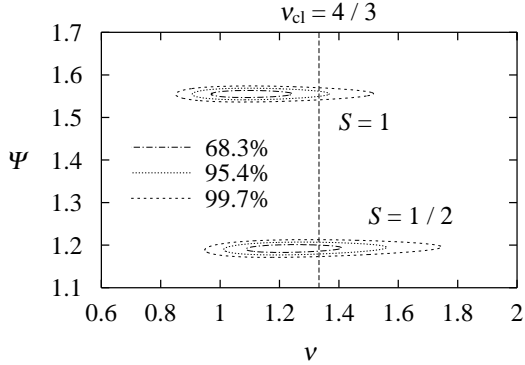}

 \caption{Confidence region of $\Psi$ and $\nu$. The percentages
 68.3$\%$ (1-$\sigma$), 95.4$\%$ (2-$\sigma$),
and 99.7$\%$ (3-$\sigma$) represent the probability that the
true parameter values fall within the confidence regions.
The classical value of $\Psi$ is 1.79167. The vertical dashed line indicates the classical value of $\nu$, 4/3.}
\end{figure}

\begin{table}[b]
\caption{Summary of critical exponents $\Psi$, $\nu$, $\alpha$, and
 $\beta$ for $S=1/2$, $1$ and the classical case ($S=\infty$). The
 values of $\Psi$ and $\nu$ are
 obtained by the FSS analysis shown in Figs. 1(a) and (b), and $\alpha$ is calculated 
 as $\alpha=2D-d-\Psi$, and $\beta$ as $\beta=(d-\Psi)\nu/2$.}

\label{table:1}
\begin{tabular}{cllll}
    $S$  & $~~~$ $\Psi$ & $~~~$ $\nu$ & $~~$ $\alpha$ & $~~~$ $\beta$ \\ \hline
    1/2  & 1.192(9)& 1.23(16) &  0.600(9) & 0.50(7)   \\
     1   & 1.555(8)& 1.09(14) &  0.237(8) & 0.24(4)   \\ \hline
 $\infty$& 1.79167 & 1.33333  &  0.       & 0.13889   \\
\end{tabular}
\end{table}

The value of $\Psi$ for $S=1/2$ obtained above is consistent with the
one ($=1.17(6)$) that has been estimated by Kato {\it et al}. from
$S_{\rm s}(L,0,p)$ at $p=p_{\rm cl}$. In their 
analysis $S_{\rm s}(L,0,p_{\rm cl})$ is approximated by 
$S_{\rm s}(L,T,p_{\rm cl})$ at low temperatures where its $T$-dependence 
becomes not discernible within the error bars. Kato {\it et al}. have also
performed the FSS analysis making use of the data at
all temperatures they have simulated. This analysis yields 
$\Psi=1.27(2)$ which differs distinctly from the present result listed
in Table I. This discrepancy may be due to the systematic error in our estimates where $S_{\rm s}(L,0,p)$ is approximated by $S_{\rm s}(L,T,p)$ at a small but finite temperature as described before.  
It should be remedied when the finite-temperature FSS
analysis is carried out as done by Kato {\it et al}. 

From Fig. 2 one sees that the accuracy of $\nu$ is significantly 
poorer than that of $\Psi$, which is interpreted as follows. The value
of $\Psi$ can be essentially extracted solely from 
$S_{\rm s}(L,0,p_{\rm cl})$, while to 
evaluate $\nu$ we need $S_{\rm s}(L,0,p)$ at $p$ other than 
$p_{\rm cl}$, or  $\delta S_{\rm s}(L,0,p)/\delta p$ at 
$p=p_{\rm cl}$. Naturally, the statistical error of $\nu$ is
expected to be larger than that of $\Psi$. 
Concerned with the systematic error in $\nu$ arising from finite-temperature
corrections, on the other hand, the saturation temperature of $S_{\rm s}(L,T,p)$ becomes smaller for smaller $|p-p_{\rm cl}|$, i.e., the effect of
finite-temperature corrections to $S_{\rm s}(L,T,p)$ becomes maximal exactly 
at $p=p_{\rm cl}$. This implies that the leading order
finite-temperature correction to $\delta S_{\rm s}(L,0,p)/\delta p$ at
$p=p_{\rm cl}$ disappears, and so the systematic error in $\nu$ is
much smaller than that in $\Psi$.

\begin{figure}[tb]
 \epsfxsize=0.47\textwidth
 \epsfbox{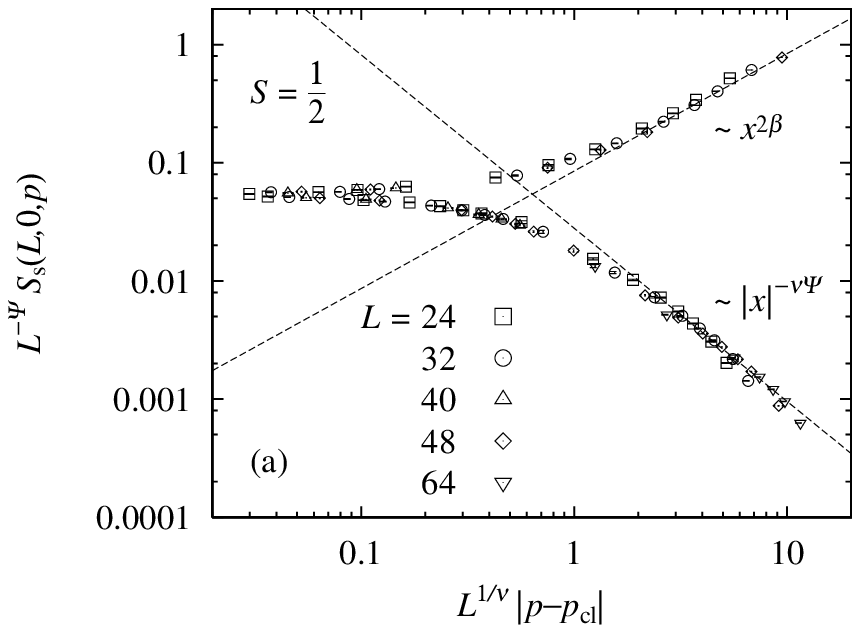}
 \vspace*{0.5em}\hspace*{.0em}

 \epsfxsize=0.47\textwidth
 \epsfbox{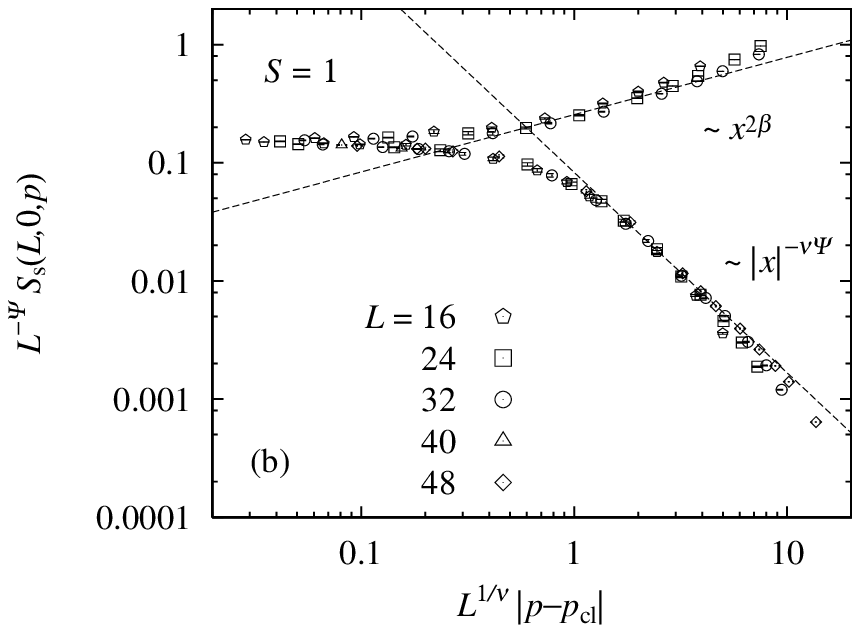}
 \vspace*{0.5em}\hspace*{.0em}

 \caption{The double-logarithmic plot of $L^{-\Psi}S_{\rm s}(L,0,p)$
 against $L^{1/\nu}|p-p_{\rm cl}|$ for (a)
 $S=1/2$ and (b) $S=1$. Dashed lines represent $ax^{2\beta}$ and
 $b|x|^{-\nu\Psi}$.}
\end{figure}

Next let us discuss $S_{\rm s}(L,0,p)$ in the full range of $p$ we have
simulated, i.e., $0.2\leq p \leq 1$, which is much wider than that in
Fig. 1. As shown in Figs. 3(a) and (b) respectively for $S=1/2$ and
1, all QMC data turn out to lie on a universal curve when 
$L^{-\Psi}S_{\rm s}(L,0,p)$ are plotted  against 
$L^{1/\nu}|p-p_{\rm cl}|$ by using the exponents $\Psi$ and $\nu$ listed
in Table I. For $L^{1/\nu}|p-p_{\rm cl}| > 1$,
the data points merge to the dashed line, which represents $ax^{2\beta}$ 
for $p>p_{\rm cl}$ and $b|x|^{-\nu\Psi}$ for $p<p_{\rm cl}$, where
$\beta$, $\nu$, and $\Psi$ are those listed in Table I, and $a$ and $b$
are arbitrary constants adjusted to fix the position of the asymptotic 
lines. At much larger $L^{1/\nu}|p-p_{\rm cl}|$ the scaling fit becomes
deteriorated, indicating that the corresponding $p$ is out of the
scaling region. The results shown in Fig. 3 also support the
scenario due to Kato {\it et al}. on the non-universal quantum phase
transition. 

In summary, in order to establish nature of the non-universal
quantum phase transition of 2D site-diluted HAF's for $S=1/2$ and $1$,
we have performed the QMC simulation in a relatively wider region of
concentration than that of Kato {\it et al}. and have estimated the critical
exponent $\nu$ more systematically. We have observed that the static
staggered structure factor is well described by the scaling form of 
Eq. (\ref{fss}) with Eq. (\ref{sf}). In particular, the exponent $\nu$
is confirmed to coincide with the classical one. These results support
the following arguments by Kato {\it et al}.: 1) there
exists no other macroscopic characteristic lengths than $\lambda(p)$,
the mean size of connected spin clusters at concentration $p$
and 2) the staggered spin correlation function between two sites on a
fractal cluster decays in a power law as 
$C(i,j;p) \sim r_{i,j}^{-\alpha}$, where $\alpha$ depends on the
strength of quantum fluctuations specified by the spin size $S$.

Most of numerical calculations for the present work have been performed
on the CP-PACS at University of Tsukuba, Hitachi SR-2201 at
Supercomputer Center, University of Tokyo, and SGI2800 at Institute for
Solid State
Physics, University of Tokyo. The present work is supported by the ``Large-scale Numerical Simulation Program'' of Center for Computational Physics, University of Tsukuba, and also by the ``Research for the Future Program'' (JSPS-RFTF97P01103) of Japan Society for the Promotion of Science.

\end{document}